\documentclass[vecphys]{article}

\usepackage{makeidx}         
\usepackage{graphicx}        
\usepackage{multicol}        
\usepackage{cite}            
\usepackage[bottom]{footmisc}
\usepackage{epstopdf}
\usepackage{lscape}
\usepackage{amsmath}
\usepackage{float}
\usepackage{authblk}

\makeindex             

\begin{document}

\title{Design Principles for Plasmonic Nanoparticle Devices}
\author[1, 2]{Phillip Manley}
\author[2]{Sven Burger}
\author[2]{Frank Schmidt}
\author[1]{Martina Schmid}
\affil{Helmholtz Zentrum Berlin f\"ur Materialien und Energie GmbH,
Hahn-Meitner-Platz 1,
14109 Berlin,
Germany}
\affil{Zuse Institute Berlin,
Takustrasse 7,
14195 Berlin,
Germany}

\maketitle

\begin{abstract}
For all applications of plasmonics to technology it is required to tailor the resonance to the optical system in question. This chapter gives an understanding of the design considerations for nanoparticles needed to tune the resonance. First the basic concepts of plasmonics are reviewed with a focus on the physics of nanoparticles. An introduction to the finite element method is given with emphasis on the suitability of the method to nanoplasmonic device simulation. The effects of nanoparticle shape on the spectral position and lineshape of the plasmonic resonance are discussed including retardation and surface curvature effects. The most technologically important plasmonic materials are assessed for device applicability and the importance of substrates in light scattering is explained. Finally the application of plasmonic nanoparticles to photovoltaic devices is discussed.
\end{abstract}

\noindent
\section{Plasmonic Principles}
\label{Plasmonic Principles}
\begin{figure}[h!]
\centering
\includegraphics*[width=.4\textwidth]{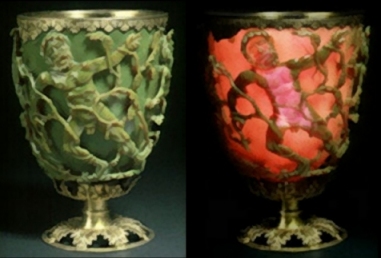}
\caption[]{The Lycurgus cup has colloidal metal nanoparticles suspended in glass. Due to plasmonic effects the glass appears green when viewed in reflection and red in transmission. \copyright Trustees of the British Museum}
\label{Lycurgus_Cup}       
\end{figure}

The most interesting properties of metal nanoparticles can be observed in the example of the Lycurgus cup\index{Lycurgus Cup}. This is a piece of late antique art currently held by the British Museum. The cup is made of metal artwork interconnected with glass, as shown in Fig. \ref{Lycurgus_Cup}. This special glass when, illuminated from the outside i.e. viewed in reflection, has a jade coloring, however when illuminated from inside, i.e. in transmission, the glass has a crimson color. This most curious effect is due to the glass containing an extremely dilute distribution of metal nanoparticles, roughly 300 parts per million \cite{LycurgusCup}. On the basis of this fact alone, what properties of metal nanoparticles can we infer? The particles must have a strong response when illuminated with light, otherwise such a dilute distribution of particles would not have such a visible effect. Additionally the effect must be wavelength dependent to generate different colors in reflection and transmission. The rest of this section will introduce plasmonics and with the aim in mind of explaining the processes occuring when light shines on the Lycurgus cup.

\subsection{Bulk Plasmons}

\begin{figure}[t]
\centering
\includegraphics*[width=.7\textwidth]{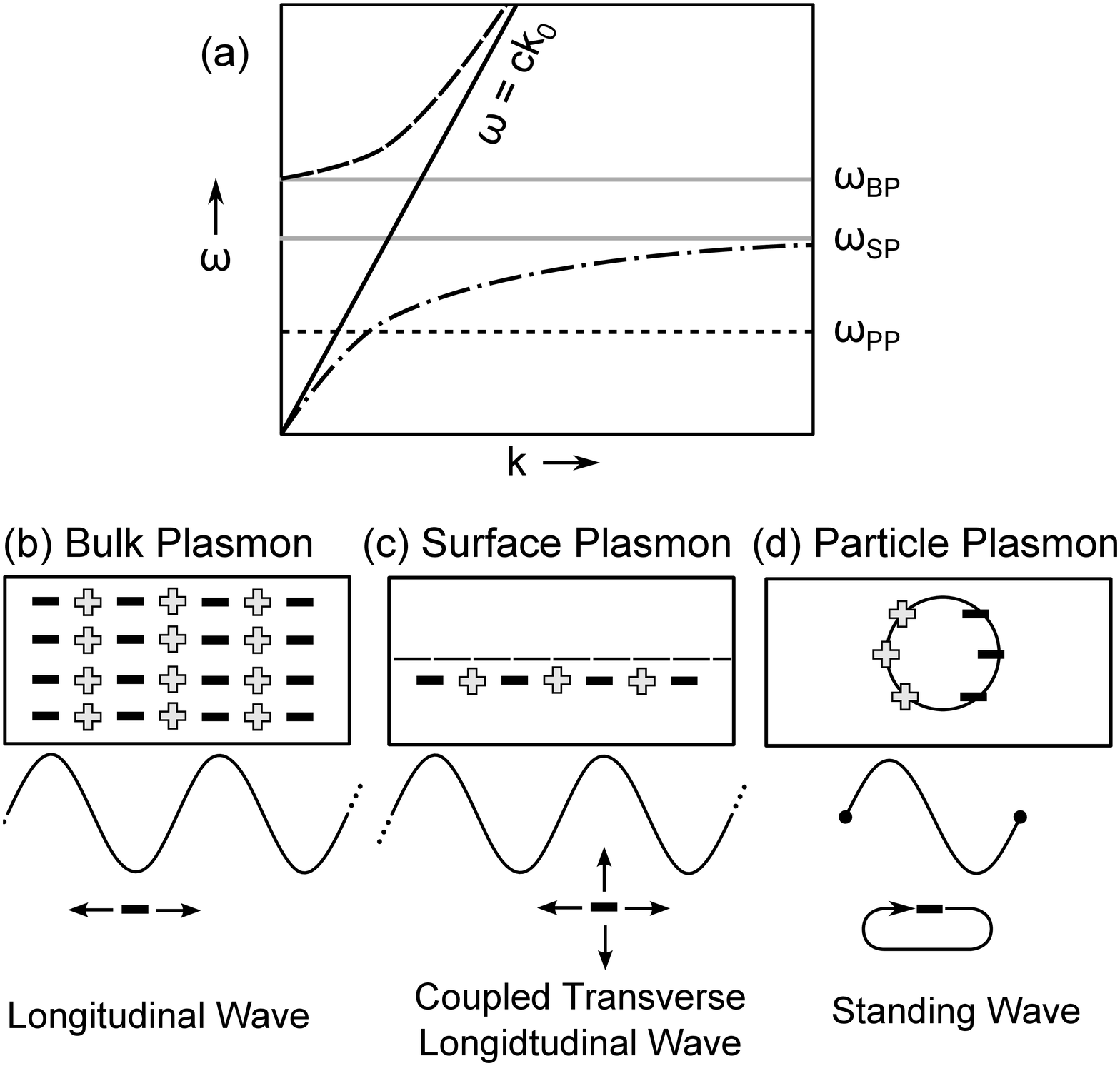}
\caption[]{(\textbf{a}) The dispersion for light in vacuum (\textit{solid line}), light in a bulk metal (\textit{dashed line}), surface plasmon polariton at a metal vacuum interface (\textit{dashed dotted line}) and particle plasmon (\textit{dotted line}). The three kinds of plasmonic resonance: (\textbf{b}) the bulk plasmon resonance occurs at $\omega_{BP}$ and is a longitudinal wave. (\textbf{c}) The surface plasmon polariton is a coupled transverse longitudinal wave bound to a metal/dielectric interface.  (\textbf{d}) The particle plasmon which is a transverse standing wave}
\label{Different_Plasmons}       
\end{figure}

A plasma\index{Plasma} is a kind of fluid consisting of positively and negatively charged particles. A plasmon is a collective oscillation of the charge density of a plasma. The free electrons and ion cores in a metal constitute a plasma with a net neutral charge. Local deviations in electron density will cause restoring forces from the fixed ionic cores leading to simple harmonic motion. The frequency of these oscillations can be easily derived \cite{Kittel} and is given by the bulk plasma frequency\index{Plasma Frequency},
\begin{equation}
\label{Plasma_Frequency}
\omega_{BP} = \sqrt{\frac{n e^{2}}{m e_{0}}}\;.
\end{equation}
Where $n$ is the free electon density in the solid, $e$ is the electronic charge, $m$ is the effective mass of the free electrons and $\epsilon_{0}$ is the permittivity of free space. The electromagnetic field distribution corresponding to a plasmonic excitation corresponds to a solution of the linear Maxwell's equations. The equations allow for both longitudinal and transverse solutions. The longitudinal solutions are associated with charge density oscillations. The transverse solutions are associated with the propagation of radiation in the medium. In the absence of an external magnetic field the two solutions are not coupled together meaning that light cannot excite charge density oscillations. Figure \ref{Different_Plasmons}(b) shows a sketch of the electronic density oscillations\index{Electronic Density Oscillations}. In the upper part of Fig. \ref{Different_Plasmons}(a), $\omega_{BP}$ is marked by a grey horizontal line, this is merely a guide to the eye and does not represent a dispersion curve. The bulk plasma frequency for most metals is higher than the frequency of visible light. When light is incident at frequencies lower than the plasma frequency, the incident light causes the electrons to move in such a way that they emit an electric field which cancels out the incident wave and creates a reflected wave. Above $\omega_{BP}$ the electrons cannot keep up with light oscillations and light will be transmitted through the sample. In the upper part of Fig. \ref{Different_Plasmons}(a) there are two lines, the solid black line shows the dispersion of light in vacuum where $\omega = ck_{0}$, the dashed line shows the dispersion of the propagation of light inside the metal which is given by \cite{Kittel},
\begin{equation}
\label{Bulk_Dispersion}
\omega^{2} = \omega^{2}_{p} + c^{2}k^{2}\;,
\end{equation}
where $c$ is the speed of light in vacuum.  For $ck >> \omega_{p}$ this converges to the free space solution. Due to light not being coupled to this kind of plasmon is not of technological importance for photonic devices.
\subsection{Surface Plasmon Polaritons}
There is a second kind of electron density wave which will have coupled transverse and longitudinal components along the surface of a metal. This is schematically shown in Fig. \ref{Different_Plasmons}(c) \cite{Surface_Plasmon_Review2}\cite{Surface_Plasmon_Review}. This happens at a lower frequency than $\omega_{BP}$,  the upper limit is shown in Fig. \ref{Different_Plasmons}(a) as grey horizontal line at $\omega_{SP}$. Since this type of wave has coupled longitudinal and transverse parts we call it a surface plasmon polariton\index{Surface Plasmon Polariton} (SPP). The dispersion relation\index{Dispersion Relation} for a SPP can be derived using Maxwell's equations \cite{Plasmonics_Reivew_Maier} and is given by,
\begin{equation}
\label{Surface_Dispersion}
k = \frac{\omega}{c}\left( \frac{\epsilon_{1}\epsilon_{2}}{\epsilon_{1}+\epsilon_{2}} \right)\;,
\end{equation}
where $\epsilon_{1}$ and $\epsilon_{2}$ are the permittivities of the metal and dielectric either side of the surface. In the limit of large $k$ and assuming the dielectric is vacuum and a Drude model for the metallic permittivity we obtain,
\begin{equation}
\omega_{SP} = \frac{\omega_{BP}}{\sqrt{2}}\;.
\label{SP_Resonance_Frequency}
\end{equation}
The well known difficulty in exciting an SP is that there is a momentum mismatch between light and surface plasmons in the same medium which means they will not be excited by light under normal circumstances. This is demonstrated in Fig. \ref{Different_Plasmons}(a), again we note that the solid line shows the dispersion of light in a vacuum, whilst the dash dotted line gives the dispersion of a SPP at a vacuum metal interface. The SPP dispersion is close to the light line in vacuum dispersion for long wavelengths. We say that the SPP is lightlike and in this regime the transverse components will dominate. For shorter wavelengths the SPP dispersion deviates significantly from the light dispersion. Here the longitudinal modes dominate and the SPP is named plasmon like \cite{Surface_Plasmon_Review2}. As we can see the SSP line has a higher momentum (larger $k$) than the light line for all frequencies. Thus we cannot conserve momentum and energy to convert light into a surface plasmon polariton. This can be circumvented by the use of a prism to increase the light momentum and evanescently couple to the surface \cite{Otto_Configuration} \cite{Kretschmann_Configuration}, or by creating a periodic structure in the metallic surface which gives momentum to the light \cite{Periodic_Grating}.

\subsection{Particle Plasmons}
The third kind of plasma oscillation shown in Fig. \ref{Different_Plasmons}(d) is the particle plasmon\index{Particle Plasmon} also called the local surface plasmon resonance\index{Local Surface Plasmon Resonance}. This is a standing wave of the electron density inside the particle, with surface charge acting as the restoring force. Due to the restoring force being the surface charge and not simply the entire fixed ion lattice as with the bulk plasmon, the particle plasmon is a transverse standing wave and can therefore be excited by incident light. There are no momentum matching conditions as with the surface plasmon polariton because particle plasmons are standing waves with a net momentum of zero. We might think that the light now has too much momentum, however when the light is absorbed by a conduction band electron and momentum is transferred to the electron, it can be easily dissipated by phonons. In other words the radiation pressure of the incoming light does not push all the electrons out of the nanoparticle because they are held inside the particle by the much stronger force of the lattice. The particle plasmon resonance frequency depends on the polarization\index{Polarization} ($\alpha$) in the particle, which for a sphere is given by \cite{VanDeHulst1981},
\begin{equation}
\label{Particle_Polarisation}
\alpha = 4\pi\epsilon_{0} r^{3} \left( \frac{\epsilon_{m} - \epsilon_{d}}{\epsilon_{m} + 2\epsilon_{d}} \right)\;.
\end{equation}
Where $r$ is the sphere radius and $\epsilon_{m}$ and $\epsilon_{d}$ are the permittivities of the metal and dielectric respectively. When the denominator of this equation is equal to zero there will be a resonance. If we assume a simple Drude form for the metallic permittivity and that the surrounding dielectric is vacuum, then we reach the result,
\begin{equation}
\label{PP_Resonance_Frequency}
\omega_{PP} = \frac{\omega_{BP}}{\sqrt{3}}\;,
\end{equation}
which has been used to sketch out the position of $\omega_{PP}$ in Fig. \ref{Different_Plasmons}(a).  In this case the dotted horizontal line shows the dispersion for a particle plasmon, the light line (solid line) will be able to cross this line for any value of $k$, this means there is no momentum matching condition for particle plasmons.

It is important to know when these results can be applied and for that we need to have a firm concept of what is meant by particle\index{Particle}. Particles are objects in which all of the dimensions are small compared to the wavelength of light. For particles of sizes the same order as the wavelength, the previous results can still be applied, however we need to take into account effects like retardation which will be covered in Sect. \ref{Shape_Effects}. An object with dimensions much larger than the wavelength is for our purposes a bulk object. We should shortly mention that we only consider particles larger than 10\,nm in diameter, this is because for particles smaller than this strong quantum effects come into play and classical descriptions for the material dependent parameters in Maxwell's equations are no longer valid.

When the particle is small compared to the wavelength the phase of the electric field will not change considerably over the nanoparticle. This means that all the free electrons\index{Free Electrons} in the particle feel the same driving force, causing them to act coherently. This coherence is what causes the large optical response. The approximation which can be made to consider such a small particle is that the electric field is (spatially) constant over the whole nanoparticle. This is often called the dipole approximation\index{Dipole Approximation} as only dipolar modes will be supported by the particle if the electric field is truly constant.

It will simplify our discussion later to consider the particle plasmon system as a damped driven harmonic oscillator\index{Damped Driven Harmonic Oscillator}. The driving force ($F_{0}$) is the incident electric field, the restoring force is the surface polarization charge, which we can imagine as a spring of constant $k$ pulling the electrons back to equilibrium. The damping\index{Damping} term $\gamma$ is equivalent to the three possible decay channels of the plasmon. Firstly it can decay radiatively, emitting a photon of the same frequency as the incoming light. Secondly it can decay due to processes inside the particle, such as phonon scattering, surface scattering or defect scattering. Thirdly, if the particle is surrounded by an absorbing medium, the plasmon can directly excite an electron hole pair in the medium.
The equation of motion and its solution are given by \cite{Feynmann_Lectures},
\begin{align}
\frac{d^{2}x(t)}{dt^{2}} &= -\omega^{2}_{0} x(t) - 2\gamma \frac{dx(t)}{dt} + F_{0} e^{i\omega t}\;, \label{EOM_DDHO}\\
x(t) &= \frac{F_{0}}{2m\omega_{0}} \frac{1}{\sqrt{(\omega_{0}-\omega)^{2}+(\gamma)^{2}}}\;, \label{Amplitude_DDHO}\\
\omega_{0} &= \sqrt{\frac{k}{m}} \label{Frequency_DDHO}.
\end{align}
The term $\omega_{0}$ is the resonance frequency which according to (\ref{Particle_Polarisation}) will be the frequency at which $\epsilon_{m} = -2\epsilon_{d}$. However for our toy model we can consider $\omega_{0}$ to be determined by the spring constant $k$ of the restoring force in (\ref{Frequency_DDHO}). There will be a peak in $x$ when the denominator of (\ref{Amplitude_DDHO}) becomes small, i.e when $\omega = \omega_{0}$, and the height of this peak will be determined by the value of $\omega_{0}$ and $\gamma$. Since $\omega_{0} \propto \sqrt{k}$ a weaker restoring force leads to a lower frequency of oscillation (redshift).

Now it is clear why the optical response of metallic nanoparticles is spectrally responsive. The resonance frequency of the metallic alloy in the Lycurgus cup was at the blue end of the spectrum. This means that when the nanoparticles were irradiated with white light, the blue light was scattered more strongly. This causes a blue dominated reflection (perceived as green by the human eye) because the only light we see is the backscattered light. It also causes the red dominated transmission because although the scattered in all directions, the ratio of red to blue light becomes red dominated in the forward direction.


\subsection{Far Field Effects}
In the example of the Lycurgus cup we discussed the effects of metal nanoparticles as viewed from a large distance (compared to the wavelength). These are called far field\index{Far Field} effects and can be utilized in many optoelectronic devices such as the increase of photocurrent in photodetectors\index{Photodetectors} \cite{photodetectors}, improved outcoupling in silicon LEDs\index{LED} \cite{SiliconLED}, anti-reflection coatings \cite{AntiReflection} and increased absorption of light in photovoltaics \cite{Design_Principles}\cite{Atwater_Polman}.

\begin{figure}[t]
\centering
\includegraphics*[width=.7\textwidth]{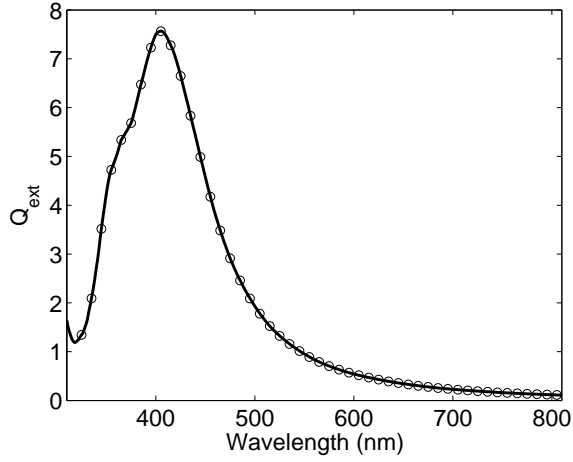}
\caption[]{The extinction cross section for a radius 50\,nm Ag nanoparticle in vacuum. (\textit{Circles}) are FEM simulation, (\textit{solid line}) is Mie theory\index{Mie Theory} simulation}
\label{Comparison_To_Mie}    
\end{figure}

All these applications rely on the unique light scattering properties of nanoparticles, therefore we now discuss the quantitative measures we can use to evaluate this light scattering. The most fundamental optical measurement of interest is to shine light at the particles and measure how much is transmitted. This results in the measurement of the spectrum of the extinction cross section\index{Cross Section} $C_{ext}$. The light can be hindered from reaching it's original destination via two different processes, it can either be absorbed in the particle or scattered by the particle into a different angle. To understand these two processes quantitatively, the total scattered ($C_{sca}$) or absorbed ($C_{abs}$) flux can be calculated using e.g. the finite element method (FEM). For some special cases, such as a single sphere, these can also be calculated with the analytical Mie theory \cite{Mie_Theory} and a comparison in Fig. \ref{Comparison_To_Mie} shows that the two methods are in complete agreement for this case. The quantity plotted is the \textit{normalized} extinction cross section \index{Normalised Cross Section}($Q_{ext}$), sometimes called extinction efficiency although the evocation of the word efficiency is in this case somewhat misleading as $Q_{ext}$ can be greater than one.


The reason we normalize $C$ is that it is important to evaluate the relative scattering strength of different sized particles fairly. The normalization is given by \cite{Bohren},
\begin{equation}
\label{NSCS}
Q = \frac{C_{Rad}}{C_{Geom}} , C_{Rad} = \frac{\Phi}{\Phi_{Inc}} , C_{Geom} = \frac{A_{Nanop.}}{A_{Comp. Dom.}}
\end{equation}
$Q$ is essentially the ratio of two terms: first the fraction of energy scattered or absorbed, which is given by the total flux scattered or absorbed $\Phi$ divided by the total flux incident to the computation $\Phi_{Inc}$; this is the cross section we have already discussed, here referred to for clarity as $C_{Rad}$. The second term is the fraction of light which would be scattered in the ray optics regime and is given by the cross sectional area of the nanoparticle $A_{Nanop.}$ divided by the cross sectional area of the computational domain $A_{Comp. Dom.}$. Concerning far field effects, in addition to the total scattering cross section we can also look at the angular distribution formed which can be crucial for certain devices designs, this topic will be explored further in Sect. \ref{Material_Effects}.



\subsection{Near Field Effects}
\begin{figure}[h!]
\centering
\includegraphics*[width=1.1\textwidth]{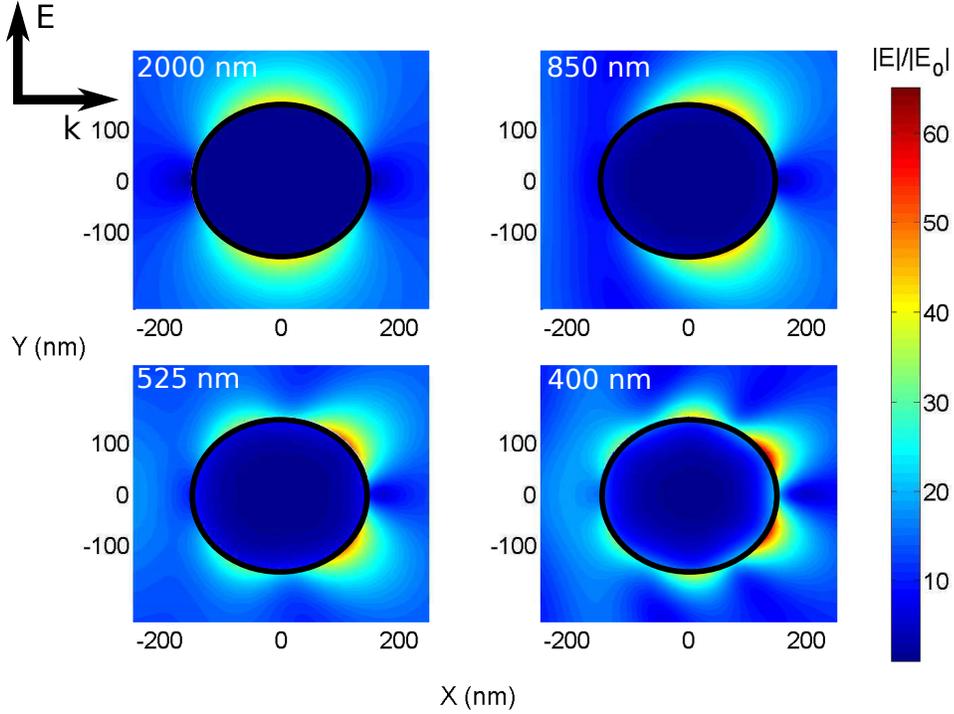}
\caption[]{The electric field intensity around a radius 150\,nm Ag nanoparticle at different wavelengths of incident light. The wavelengths are shown in the upper left corner of each box. For a wavelength of 2000\,nm the field is constant over the whole nanoparticle and the dipole field is placed at the center of the particle. At 850\,nm there will be a non negligible variation in the field over the nanoparticle leading to the dipole radiation to be asymmetrically distributed. For even shorter wavelengths the quadrupole\index{Quadrupole} and octupole mode become dominant}
\label{Multipoles}    
\end{figure}
The other way in which nanoparticles can be used in devices is to harness the strong near fields\index{Near Field} generated by the plasmon before it decays. There is an exponentially decreasing electric field extending out from the surface of the nanoparticle. The peak value of this electric field can be extremely high which opens the door to different device capabilities. The possible device applications for this strong localized electric field are numerous, therefore the following list is not exhaustive. Increasing the signal of surface enhanced Ramman spectroscopy \index{SERS}is possibly the most well known application, as the signal depends on the fourth power of the electric field, enhancement factors for $10^{15}$ have been reported \cite{SERS} \cite{SMSERS} for molecules in the vicinity of specially designed metal nanoparticles. Other applications include optical waveguides of sizes below the diffraction limit \cite{Subwavelength_Waveguide} \cite{Subwavelength_Waveguide2}, increased absorption in organic solar cells \cite{Organic_Enhancement1}\cite{Organic_Enhancement2}, biosensing \cite{Plasmonic_Biosensing}, delivery of dyes for medical applications \cite{Dye_Delivery} and separation of chiral molecules \cite{Chiral_Separation}.

 The strength of the electric field in the vicinity of the particle can also be calculated via the FEM and examples are shown in Figs. \ref{Multipoles} and \ref{Near_Field}. In all the figures the field enhancement can be seen to decay exponentially away from the nanoparticle surface. Figure \ref{Multipoles} shows the near field distribution of a large Ag nanoparticle with radius 150\,nm at four different wavelengths of incident light. For incident light of 2000\,nm the dipole approximation is valid and the near field shows a dipole pattern around the center of the nanoparticle. For the shorter wavelength of 850\,nm which is approximately where the far field dipole resonance is, we still see dipole dominated radiation. However due to the electric field of the incident light not being constant over the nanoparticle volume, there is some mixing of the dipole and higher order modes. This causes the dipole mode to shift away from the center of the nanoparticle, causing the asymmetrical resonance. At 525\,nm the transition to quadrupole dominated resonance has completed with two more resonance lobes forming on the left side of the nanoparticle. For yet shorter wavelengths the quadrupole transitions into the octupole resonance. A more detailed description of effects that different particle shapes can have on the near field resonance will be given in Sect. \ref{Shape_Effects}.



\section{Finite Element Method}
\label{Finite Element Method}
Having already discussed physical characteristics of plasmonic nanoparticles and what observable quantities we wish to simulate, it is necessary to review the numerical method which will be used during this chapter.
The finite element method\index{Finite Element Method} is a procedure for solving partial differential equations by minimizing an expression made from trial solutions with variable parameters \cite{FEM1} \cite{FEM_Maxwell}. This is true of all variational problems\index{Variational Problem}, the innovations of the finite element method are primarily in the choice of trial functions.

Generally with variational solutions as in the case of Fourier decomposition, we require many trial functions to approximate the solution correctly which can be computationally costly. The finite element method uses higher order polynomial basis functions. In the simulations presented here the order is typically between one (linear) and six, such simple functions are relatively easy to handle computationally. At first glance these functions would be assumed to give a very poor approximation to the solution; the other key to this problem is the introduction of elements. The computational domain is split up into smaller elements. In each element the solution is assumed to vary slowly enough that simple polynomial functions do provide a good approximation to the solution. The solutions over each element are then reassembled with appropriate boundary conditions to provide the whole solution.

For optical problems we must solve the Maxwell equations\index{Maxwell's Equations} \cite{Jackson}. By employing the time harmonic ansatz $\mathbf{E}(t) = \mathbf{E_{0}} e^{-i\omega t}$ and assuming no external charges or currents $\rho = \mathbf{J} = 0$ we can obtain two equations for the electric field \cite{JCMwave},
\begin{align}
\label{Maxwell}
\nabla \times \mu^{-1} \nabla \times \mathbf{E} - \omega^{2}\epsilon \mathbf{E} &= 0\;, \\
\nabla \cdot \epsilon\mathbf{E} &= 0\;,
\end{align}
where $\epsilon$ and $\mu$ are the permittivity and permeability of the element in question respectively.

There are two principle ways to increase the accuracy of our discrete solution with respect to the real solution. Either the mesh upon which we are solving the equation can be refined, e.g. divide the length of every edge in the mesh by two, or the order of approximating polynomial can be increased. We would like to use the input parameters requiring the lowest amount of computational resources. Therefore it is crucial to understand how varying these two parameters affects the precision of the solutions. Furthermore when designing devices one generally requires an optimization\index{Optimization} procedure. For such a procedure to work we must be able to compare the outcomes of different simulations with only small differences in input parameters. This will lead to a correspondingly small change in the figure of merit. To compare which set of input parameters maximizes the figure of merit it is essential to know the precision to which the simulated quantities are valid.

\begin{figure}[t]
\centering
\includegraphics*[width=.7\textwidth]{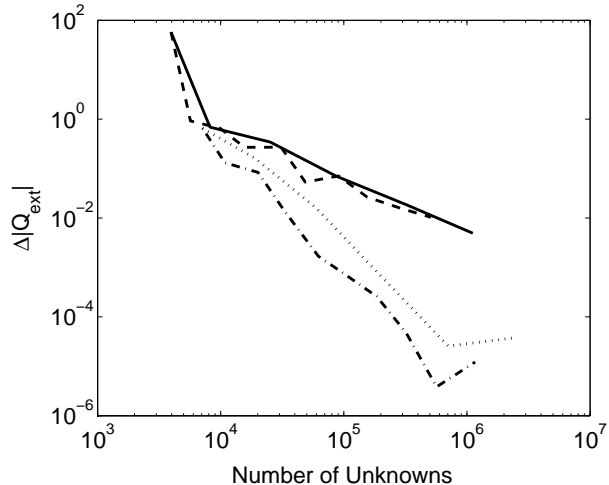}
\caption[]{The \index{Convergence}convergence of $Q_{ext}$ with respect to grid refinements. The (\textit{solid line}) and (\textit{dotted line}) show global grid refinement for finite element degree 2 and 3 respectively. (\textit{dashed line}) and (\textit{dash dot line}) show adaptive convergence for finite element degree 2 and 3 respectively; the grid is only refined in those places with a higher error until the error is equal everywhere.}
\label{Convergence_Test}    
\end{figure}
\begin{figure}
\centering
\makebox[\textwidth][c]{\includegraphics[width=1.5\textwidth]{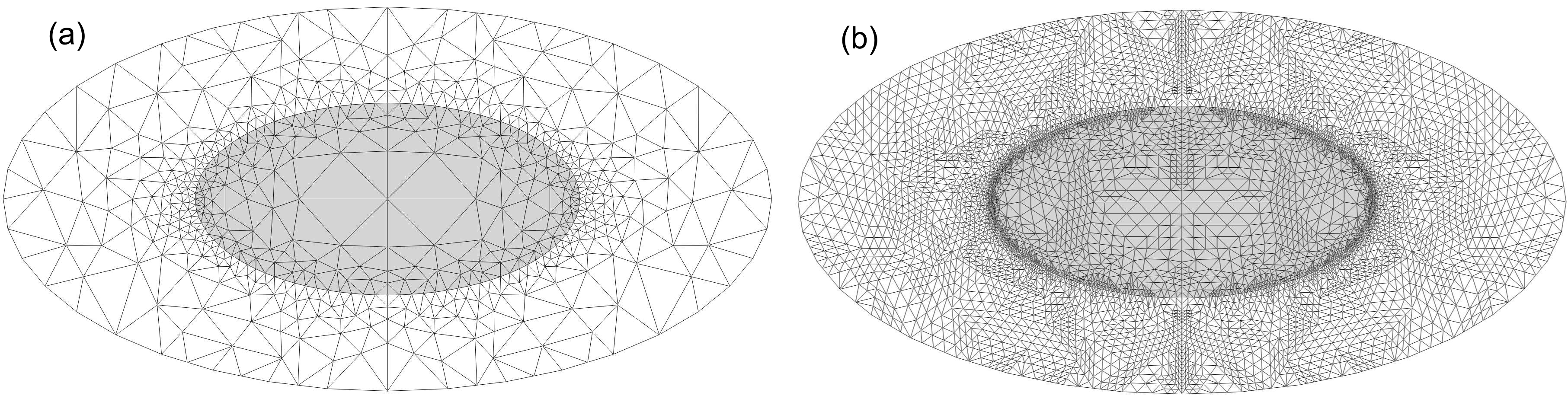}}
\caption{A cross section of the mesh grid used for a typical FEM calculation. The ellipsoidal nanoparticle had been shaded for clarity. (\textbf{a}) is an unrefined grid and (\textbf{b}) shows the same grid after three adaptive refinement steps}
\label{Grid_Refinement}
\end{figure}

Due to the method focusing on finding local solutions and then combining said solutions, we can define local error estimations\index{Adaptive Refinement}. This allows for the computational algorithm to automatically determine where the solution is least converged and locally refine the mesh there. Figure \ref{Convergence_Test} shows the convergence of $Q_{ext}$ in the case of an oblate Ag spheroid in vacuum with axes lengths of 200\,nm, 100\,nm and 100\,nm.  The computation was performed by exploiting the cylindrical symmetry of the problem. This means only a 2D representation of the particle (as seen in Fig. \ref{Grid_Refinement}) needs to be calculated, then the complete 3D solution can be reconstructed \cite{JCM_Cylindrical}. The mesh was refined using two different techniques; global refinement and adaptive local refinement. Plotted is the absolute difference between the solution with $n = 0,1,...,N$ refinement steps and the solution with $N+1$ refinement steps, the x axis shows the amount of unknowns (i.e. the computational effort required) in each case. As expected, refining the mesh increases the computational cost but also increases the precision of the solution. The method of adaptive refinement produces solutions of equal convergence for many fewer unknowns compared to the global refinement strategy.

Figure \ref{Grid_Refinement} shows a cross section of a typical mesh grid used for FEM simulations. The grid has been preferentially refined near the boundary between the nanoparticle and the surrounding medium, especially where the curvature is highest. In the later sections we will see that these correspond to areas of large electric field gradient necessitating a high local precision.

By taking the time harmonic ansatz for (\ref{Maxwell}) we have converted the problem to a frequency domain problem. This means each computation gives information for a single frequency, cf. time domain methods which can provide an entire frequency spectrum with one computation provided the time step is short enough. However each frequency domain computation is much less demanding than a time domain computation. This can be advantageous for plasmonics as the whole spectrum may not need to be sampled but just a specific subset where the plasmon resonance is located. This can be relatively cheaply computed with frequency domain methods.

Sometimes periodic arrays of nanoparticles are the object of study, in which case the choice of boundary conditions at the computational domain boundary can be naturally fulfilled by periodic boundary conditions. For the case of isolated nanoparticles we need a different kind of boundary condition. If we were to stipulate that the field should discontinuously drop to zero outside of the computational domain it would be equivalent to placing a mirror around the computational domain boundary causing reflections that have nothing to do with the physics of the problem. To eliminate these reflections we require a so called transparent boundary condition.  These are realised by an adaptive perfectly matched layer\index{Perfectly Matched Layer} (PML) method. PML is a boundary condition intended to remove these unphysical reflections \cite{PML}. It achieves this by firstly creating an extra layer outside the computational domain with the same optical constants as inside. The field in this new layer decays to zero without changing the optical constants. This is achieved by allowing the position variable outside the computational domain to become a complex variable with a steadily increasing imaginary part. The equations for traveling waves contain factors like $exp[i(\mathbf{k.r}-\omega t)]$. If $\mathbf{r}$ is allowed to become complex then we will have factors proportional to, $exp[i(\mathbf{k}.\Re\{\mathbf{r}\}-\omega t)]exp(-\mathbf{k}.\Im\{\mathbf{r}\})$, which will give an exponential decay. This complexification of the real distance is an exact extension, hence no artificial reflections will be introduced. Residual reflections are due to the discretization and get smaller with finer meshes and larger thicknesses of the PML layer. This type of boundary condition was implemented in all the simulations presented in this chapter.

\section{Shape Effects}
\label{Shape_Effects}
\begin{figure}[h!]
\centering
\includegraphics*[width=.7\textwidth]{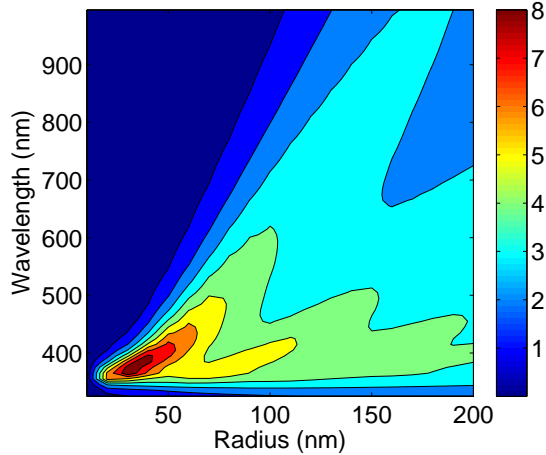}
\caption[]{Contour plot of the normalized extinction cross section for a spherical Ag nanoparticle in vacuum. Separated peaks refer to different multipole orders.  Reasons for redshifting with increasing radius are discussed in the text.}
\label{Xsec_Radius_Dependence}    
\end{figure}

In this section we will detail how the different nanoparticle shapes affect the kind of resonance observed. We begin by considering a single spherical silver particle in vacuum. The fundamental parameter is the radius of the particle, which we wish to vary and simulate the effect this has on the spectrum of $Q_{sca}$ and $Q_{abs}$. Figure \ref{Xsec_Radius_Dependence} shows a contour plot of $Q_{ext}$ with respect to the particle radius\index{Particle Radius} and the wavelength. Starting from the low radius end there is almost no response to the scattering for very small particles (10\,nm radius). A peak develops in the short wavelength regime upon increasing the radius until at 40\,nm the spectrum looks like a strong sharp peak. Upon further increasing the radius this peak redshifts, broadens and reduces in intensity. This peak is associated with the dipole mode\index{Dipole Mode} of the nanoparticle, this can be deduced from our previous reasoning that the small particles feel a spatially constant electric field which only permits a dipole mode. It can also be seen by looking at the near field as shown earlier in Fig. \ref{Multipoles}(a). For radii larger than 75\,nm we observe another peak forming at lower wavelengths than the dipole peak. This peak corresponds to the quadrupole resonance which has become significant because our particle has become so large that we can no longer consider the electric field to be spatially constant. Again by observing the near field in Fig. \ref{Multipoles} we can confirm what physical intuition would suggest; that this mode is quadrupolar. This resonance naturally occurs at a lower wavelength than the corresponding dipolar mode for a given radius. The wavelength of light required to excite the quadrupolar mode should be half of that required to excite the dipolar mode. This is because twice as many oscillations of the electric field need to fit into the save volume of nanoparticle. This mode also shows a redshifting, broadening and reduced peak intensity for increasing radius, the explanation of which is discussed below. The beginning of the octupole mode is also visible between radius 175-200\,nm. The same logic for describing the quadrupolar mode can be used here, and indeed any higher order modes which would appear at even larger radii.

All three of the \index{Multipole} orders show redshifting with increasing particle radius. When increasing the radius, one essentially changes four separate parameters. The first three are the lengths along the three principle axes, these are shown in the inset of Fig. \ref{Elongation_Dependence}. They are the \textbf{E} direction (direction of polarization), the \textbf{k} direction (direction of the wave propagation) and the \textbf{H} direction which is perpendicular to both the others. The fourth parameter is the radius of curvature of the object. To determine the relative importance of these factors we start with a sphere of radius 50\,nm, then elongate the sphere along one of the three axes whilst keeping the surface curvature constant. A sketch of the different orientations is shown in the inset of Fig. \ref{Elongation_Dependence}. Although the volume is changing in each case, the only factor contributing to the volume change is the one which we intend to investigate.

\subsection{Nanoparticle Length $\|$ to \textbf{E}}
\begin{figure}[t]
\centering
\includegraphics*[width=.7\textwidth]{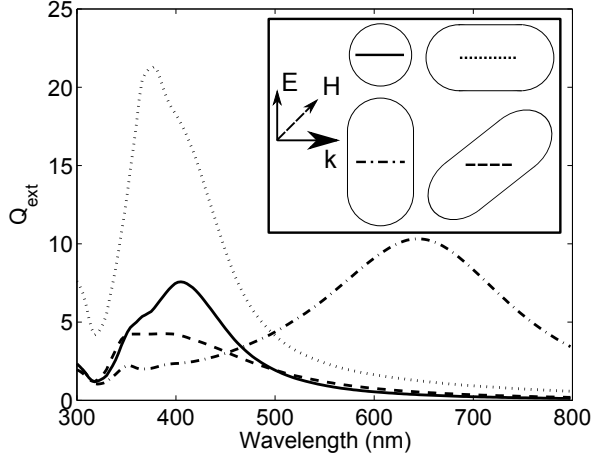}
\caption[]{The normalized extinction cross section for a radius 50\,nm Ag sphere in vacuum and three orientations of a 200\,nm long Ag cylinder in vacuum with ends having the same curvature as the sphere. The line style within each shape in the inset links the spectra to the different simulation setups}
\label{Elongation_Dependence}    
\end{figure}

\begin{figure}[t]
\centering
\includegraphics*[width=.7\textwidth]{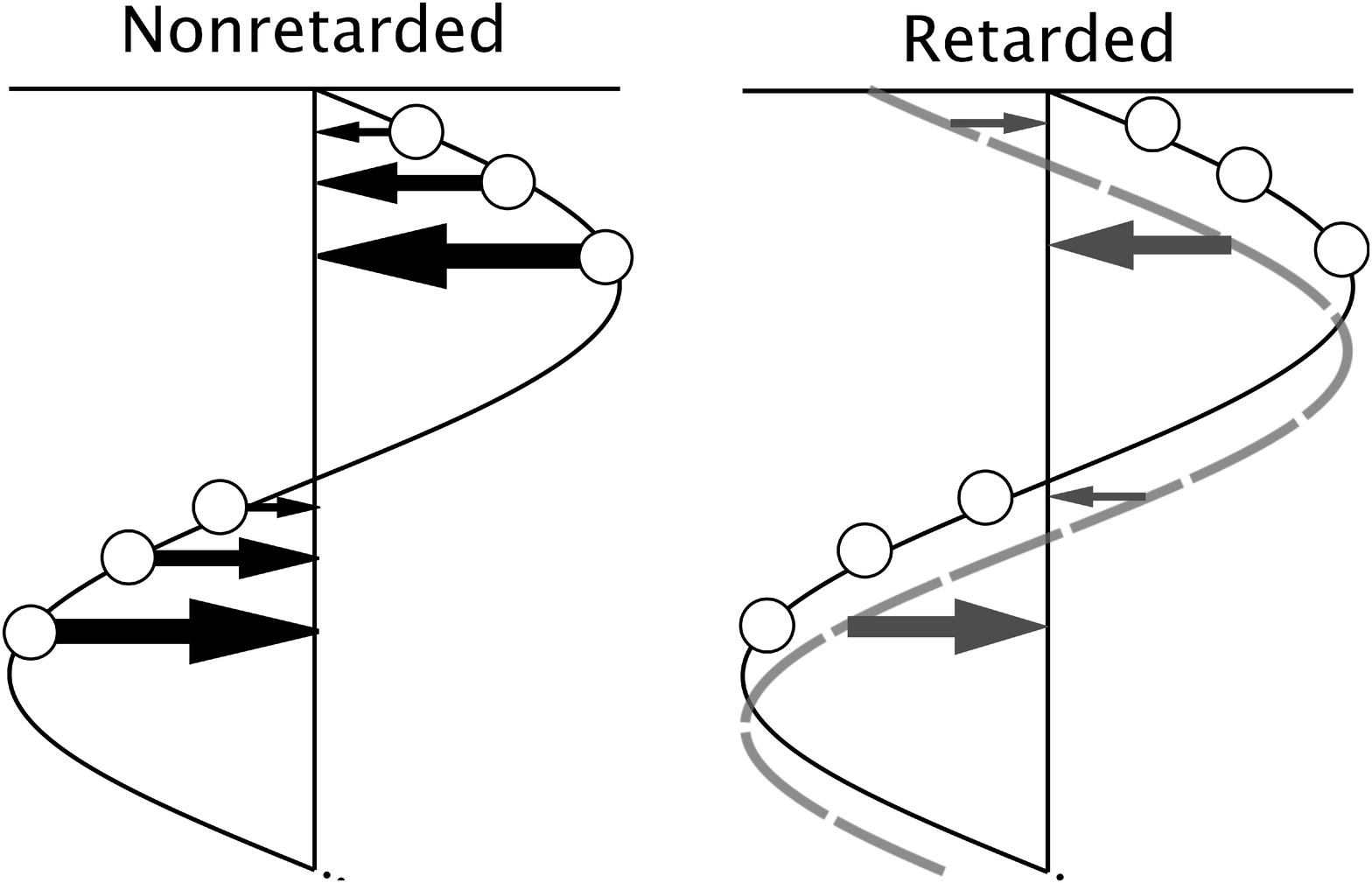}
\caption[]{Schematic view of the retarded restoring force. (\textbf{Left}) In the non retarded case the restoring force is proportional to the distance of the electron from equilibrium. (\textbf{Right}) In the retarded case the restoring force (\textit{dotted curve}) lags behind the electron's position (\textit{solid curve}) due to having to travel across the nanoparticle. This leads to a reduced restoring force compared to the non retarded case}
\label{Retardation}    
\end{figure}

We first discuss the most dominant effect which is increasing the length parallel to \textbf{E}. This will cause redshifting of the resonance due to retardation\index{Retardation} of the Coulomb force over the particle as can be seen in Fig. \ref{Elongation_Dependence}. If we were to apply a constant electric field to the nanoparticle, the clouds of electrons and ions will be shifted in opposite directions. Due to the much greater mass of the ions we can consider the acceleration felt due to the electric field to be negligible, therefore we assume the ions are static. This will give rise to a negative and positive surface charge building up causing a restoring Coulomb force to propagate from the positive charge towards the negative charge. This takes a finite time to traverse the entire nanoparticle. Now if the applied field is constant with time the difference is immaterial. However if the driving force is an electric field oscillating in time with a high enough frequency (for our discussion visible light is high enough) we have our harmonic oscillator situation. If we just consider one cycle, as in Fig. \ref{Retardation}, then as the driving force is applied the electrons start to move away from the ions. In the absence of retardation the restoring force at a given time $t_{0}$ will be of a certain strength, then at the later time $t_{1}$ the electrons have moved further away and the restoring force has become even larger, just like a spring being stretched out. This is shown in the left panel of Fig. \ref{Retardation}. Now when we consider the effect of retardation, at our first position where we measured the restoring force, we must now evaluate at the retarded time. This time, let us call it $t_{-1}$ will be in the past with respect to $t_{0}$. Thus in the non retarded case, the electron's position at $t_{-1}$ was closer to the equilibrium position, meaning the restoring force was lessened. The same holds true of the electron in the retarded restoring force at time $t_{1}$ feeling the same force as in the unretarded case at time $t_{0}$. We can now see why retardation of the Coulomb force means a weaker restoring force on the electron, which will lead to a redshift in the resonance. Quantum mechanical calculations have confirmed that this is the case \cite{Quantum_Retard}. To see that it must cause a redshift and not a blueshift take the limit of an infinitely retarded force i.e. zero restoring force. In this case the period of oscillation must be infinite i.e. no oscillation which corresponds to an infinitely large wavelength.

The spectrum is also significantly broadened as well as redshifted. This is because for longer wavelengths the radiative damping increases significantly \cite{Radiation_Damping}. This also partly explains why in Fig. \ref{Xsec_Radius_Dependence} the different multipoles reduce in intensity for longer wavelengths. The other reason being related to the reduced surface curvature for large spheres which will be elaborated upon in Subsect. \ref{Subsec:Surface_Curvature}.

\subsection{Nanoparticle Length $\|$ to \textbf{k}}
\begin{figure}[h!]
\centering
\includegraphics*[width=1.1\textwidth]{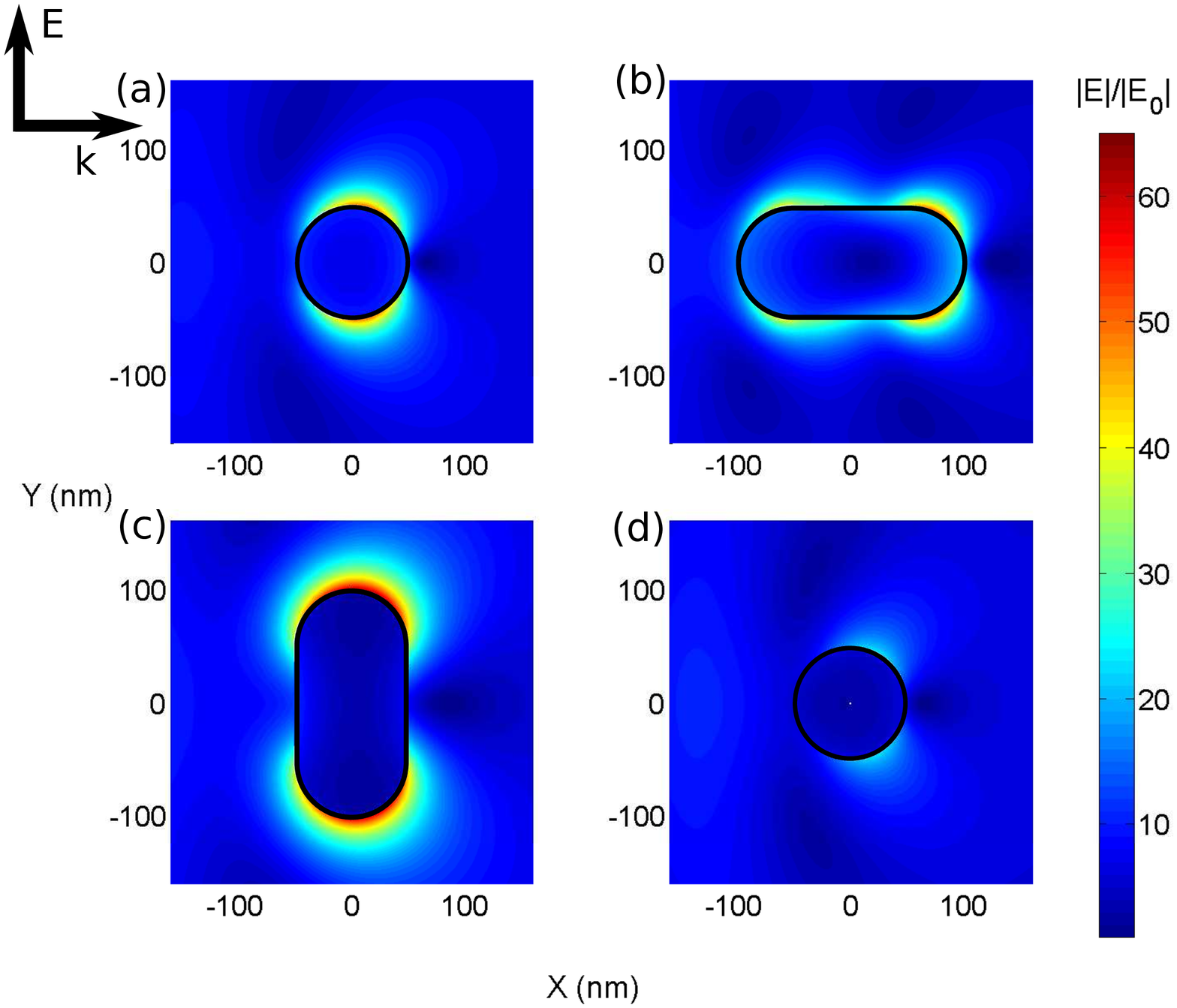}
\caption[]{The electric field intensity around different shaped nanoparticles normalized to the incident field intensity. The direction and polarization of the incident light is shown in the top left corner. Black lines have been annotated to show the particle outline. (\textbf{a}) A spherical Ag particle of radius 50\,nm in vacuum, (\textbf{b}),(\textbf{c}) and (\textbf{d}) an Ag cylinder of length 200\,nm in vacuum, with radius of curvature equal to the sphere. The orientations of the nanoparticles are the same as the inset in Fig. \ref{Elongation_Dependence}. The wavelength in each case is the peak wavelength of the spectrum in Fig. \ref{Elongation_Dependence}.  The viewing plane is cut through the center of the nanoparticle}
\label{Near_Field}    
\end{figure}

Next we discuss the effect of changing the length of the particle in the direction of wave propagation. As shown in Fig. \ref{Elongation_Dependence} this causes a slight blueshift and a large increase in intensity. The slight blueshift is due to the small reduction in surface curvature in the direction of polarization, which will be dealt with in more detail in Subsect. \ref{Subsec:Surface_Curvature}. If the surface curvature would not have changed then there would be no blueshift. The increase in intensity comes from our normalization condition, we normalized to the area of the particle as seen by the incoming light. When the light is parallel to the long axis this area does not change even though the volume increases. There has been a change in the lineshape\index{Lineshape} of the resonance. The sphere shows a peak with a shoulder on the short wavelength side signifying a smaller peak there that is not well resolved due to the larger peak. For the cylinder the case is reversed with the main peak being at the lower wavelength side with a shoulder at the longer wavelength side. This is due to the change from a dipole dominated resonance to a quadrupole dominated resonance. This effect is visible in the near field shown in Fig. \ref{Near_Field}(b) which has four nodes at the nanoparticle boundary for the cylindrical case. When only half a wavelength or less can fit inside the nanoparticle the dipole mode will dominate, whereas for particles larger than half a wavelength of light, the multipole orders begin to dominate. Therefore it is important to keep the length of the particle in the direction of wave propagation short if one wishes to avoid multipolar modes.
\subsection{Nanoparticle Length $\|$ to \textbf{H}}
Furthermore we look into changing the particle length parallel to \textbf{H}. Here we can see essentially no redshifting or blueshifting but simply a reduction in intensity of the resonance. The near field representation of this resonance in Fig. \ref{Near_Field}(d) shows a reduced intensity in the electric field enhancement around the nanoparticle in the plane of polarization. This reduction in intensity of the resonance is due to the particle loosing \index{Coherence} coherency of oscillations, i.e. the electrons in the center of the particle are not oscillating in phase with those at the edges.

\begin{figure}[t]
\centering
\includegraphics*[width=.7\textwidth]{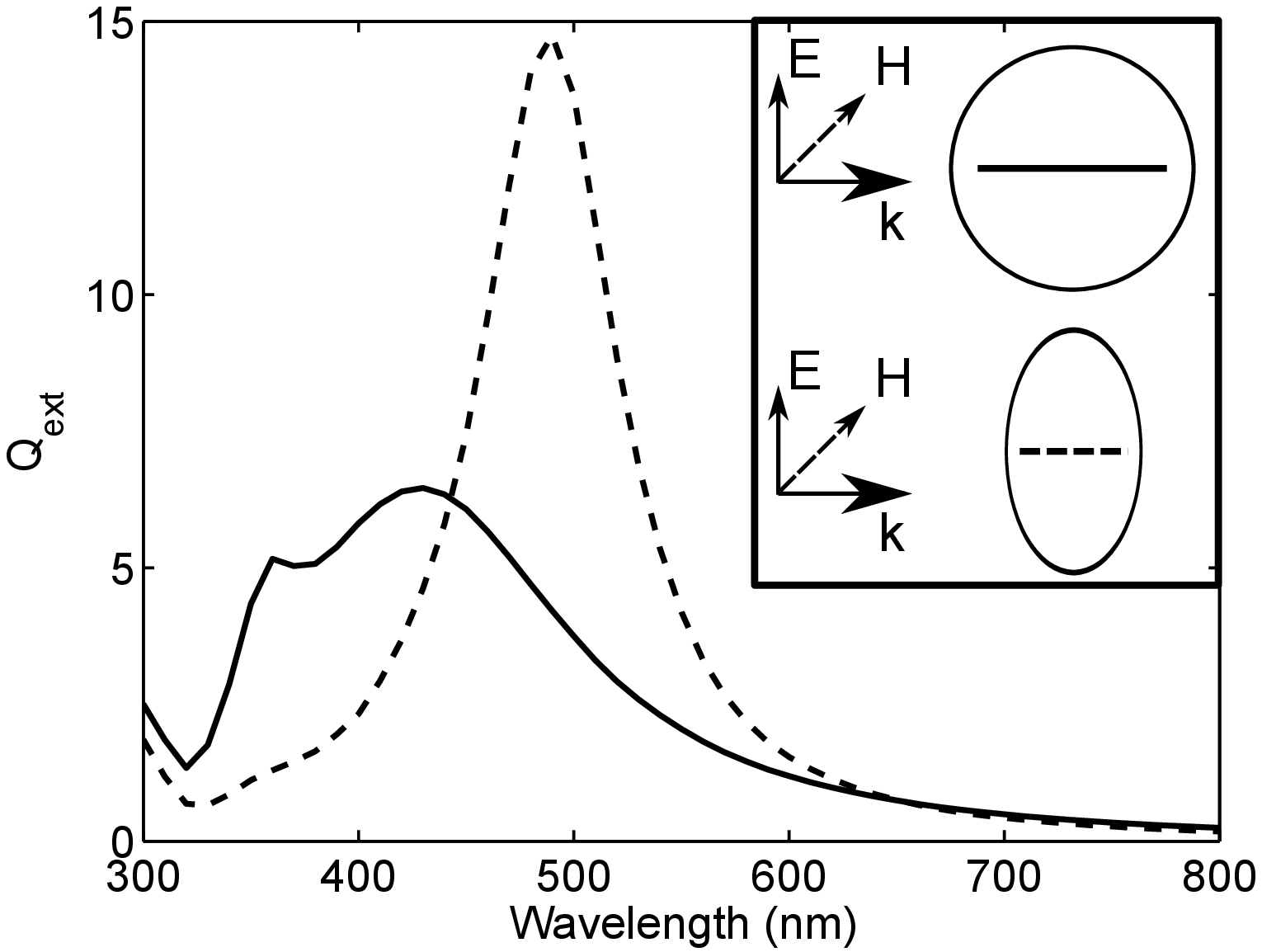}
\caption[]{The normalized extinction cross section for (\textit{solid line}) a radius 60\,nm Ag sphere in air and (\textit{dashed line}) an ellipsoid with the same radius as the sphere in the direction of polarization, but with 30\,nm radii in the other directions}
\label{Surface_Curvature}    
\end{figure}
\subsection{Surface Curvature}
\label{Subsec:Surface_Curvature}
Finally we come to the effect of surface curvature\index{Surface Curvature}. We start with a 60\,nm radius sphere and shorten two of the axes equally, thus creating a prolate ellipsoid with increased surface curvature along one axis, along which we choose to direct the polarization. Notice that in this configuration we have kept the length $\|$ to \textbf{E} constant between the two cases. This is because changing this length had the dominant effect on the spectral position of the resonance as we have seen in Fig. \ref{Elongation_Dependence}, thus any changes to the spectrum in this case can be attributed to the effect of surface curvature alone. Figure \ref{Surface_Curvature} shows the spectra for the original 60\,nm radius sphere and for ellipsoid with polarization along the 60\,nm axis. When the polarization of the incident light is along the axis which has been elongated there is a redshifting of the resonance and also an increase in intensity.

As we have already seen, redshift\index{Redshift} effects are associated with a reduced restoring force in the harmonic oscillator equations. The reduction in restoring force is due to the higher surface to volume ratio in the ellipsoid. In the case we have considered the volume of the sphere is four times that of the ellipsoid, whereas the surface area of the sphere is only $\approx 2.5$ times larger than the ellipsoid. This means that the fixed amount of volume charge is spread more thinly over the surface in the case of the ellipsoid leading to a reduced restoring force. This will naturally lead to an increased intensity of oscillation. However there is another factor which contributes to an increased near field intensity for high surface curvatures. The electric fields around sharp metallic points are stronger for all wavelengths, not just those at which plasmonic effects are present. This is due to the boundary condition that electric fields must exit a metal normal to the surface, therefore at curved surfaces there is a bunching of the field lines. This effect is commonly referred to as the ``lightning rod effect''\index{Lightning Rod Effect} and is not specific to plasmonics \cite{Lightning_Rod}.


\section{Material Effects}
\label{Material_Effects}
\subsubsection{Nanoparticle Material}
\label{Subsec:Nanoparticle_Material}
\begin{figure}[t]
\begin{center}
\makebox[\textwidth][c]{\includegraphics*[width=1.5\textwidth]{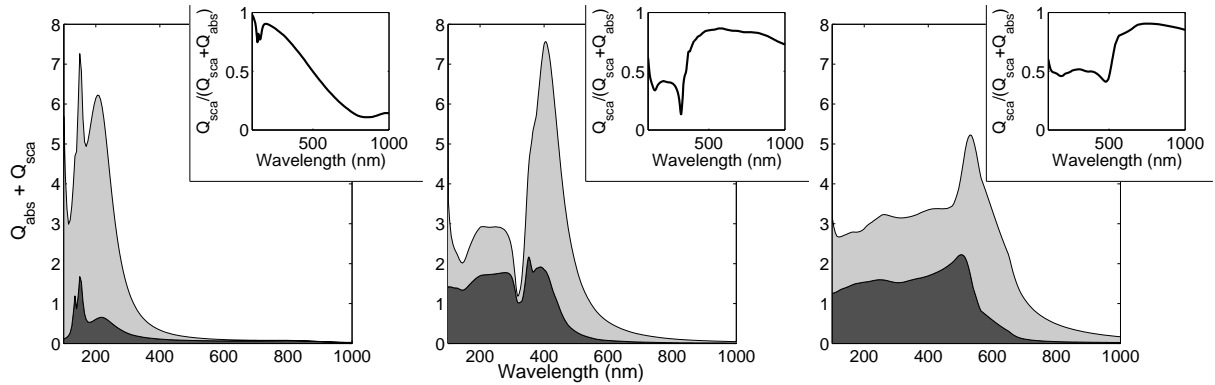}}
\caption[]{ Stacked area plot of normalized extinction cross section (\textit{decomposed into absorption - dark grey} and \textit{scattering - light grey}) for (\textbf{left}) aluminium, (\textbf{center}) silver and (\textbf{right}) gold. Inset shows the fraction of light scattered in each case. Each case shows a nanoparticle in vacuum with a radius chosen to maximise the extinction cross section, radii are Al - 30\,nm, Ag - 50\,nm and Au - 80\,nm}
\end{center}
\label{Metals_Comparison}    
\end{figure}

The optical properties of metal nanoparticles depend strongly on the metal in question. Figure \ref{Metals_Comparison} shows $Q_{ext}$ for aluminium\index{Aluminium}, silver\index{Silver} and gold\index{Gold}. $Q_{ext}$ is decomposed into the absorption and scattering. The inset shows $Q_{sca} / Q_{ext}$ for each case which we will refer to as scattering efficiency, note that in some literature $Q_{sca}$ is called the scattering efficiency. Each case has a spherical nanoparticle in vacuum with radius chosen to maximise $Q_{ext}$, i.e. 30 nm, 50 nm and 80 nm respectively. The Al plasma resonance forms two distinct peaks as the quadrupole mode has already started to become prominent at a nanoparticle radius of 30\,nm. The resonance shows a high scattering efficiency that quickly falls off with increasing wavelength. The minimum in scattering efficiency occurs at 800\,nm and is due to an absorption from an interband transition. This could prove problematic for usage of Al at optical wavelengths, nevertheless Al has a strong scattering response, comparable to Ag, and could provide a lower cost alternative. Ag has a strong scattering response and a high scattering efficiency over the whole optical region due to interband transitions\index{Interband Transitions} starting below 300\,nm. Au has a weaker response but also a high scattering efficiency. Due to the interband transitions starting below 500\,nm, Au cannot be used over the whole optical range but is a strong candidate for IR applications.

The general principles for choosing plasmonic materials\index{Plasmonic Materials} for optical or IR devices are $\epsilon^{\prime}$ negative and $\epsilon^{\prime\prime}$ as close to zero as possible in the frequency range of interest \cite{Plasmonic_Materials_Review2}. The first condition is equivalent to having the bulk plasma frequency higher than the frequency range of interest. For the optical region this condition is met by the noble metals. For the IR region this condition can be met by alkali metals, highly doped semiconductors and graphene \cite{Plasmonic_Materials_Review}. In the case of semiconductors it is also necessary for the band gap frequency to be higher than the frequency range of interest to keep the loses low.

For realistic device design more factors than just the optical properties must be taken into account. The chemical properties will also be crucial; copper and aluminium both easily form oxides\index{Oxides} which can hinder the plasmonic capabilities. Likewise alkali metals are extremely reactive therefore their use in real devices would be problematic.

\subsubsection{Homogeneous Medium}
\label{Subsec:Homogeneous_Medium}
\begin{figure}[t]
\centering
\includegraphics*[width=.7\textwidth]{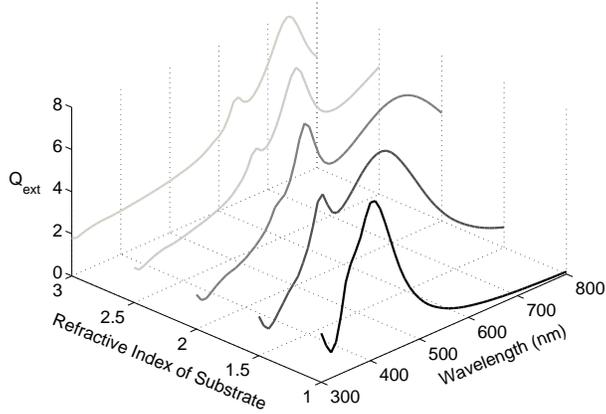}
\caption[]{ The normalized extinction cross section for a spherical Ag particle of radius 50\,nm in media with different refractive indices}
\label{Homogeneous_Index}    
\end{figure}

Another factor which can be used to control the plasma resonance is the nanoparticle's surrounding medium\index{Medium}. In situations where the particles are suspended in a colloid or embedded totally in another material or have a shell structure, the outer medium can be taken as homogeneous. In Fig. \ref{Homogeneous_Index} the extinction cross section of a radius 50\,nm Ag sphere is shown for different values of refractive index. The absorption in the surrounding medium is zero. The extinction cross section is redshifted with increasing index of the medium. This is because the surface charge density of the nanoparticle will be partially compensated by the surrounding dielectric atoms causing a weakened restoring force \cite{Plasmonics_Reivew_Maier}. As we have seen a weaker restoring force (spring constant $k$) will lead to lower resonance frequency $\omega_{0}$ for a damped driven harmonic oscillator\index{Damped Driven Harmonic Oscillator}. Different multipole modes are redshifted at different rates, thus we are able to see in Fig. \ref{Homogeneous_Index} the separation of distinct peaks which previously were one combined peak. The dipole peak is redshifted the strongest and also undergoes significant broadening until for $n>2.0$ it is shifted out of the wavelength range shown. The quadrupole peak is also strongly shifted and a third peak corresponding to octupole modes starts to emerge.

In our harmonic oscillator (\ref{Amplitude_DDHO}) example we saw that the amount of damping in the system will affect both the broadening of the resonance and the amplitude. Therefore a broader resonance will naturally have a lower amplitude. The resonance amplitude will also increase when the resonant frequency $\omega_{0}$ is redshifted. However if the damping increases at a larger rate then the net effect will be a redshift with no increase in amplitude. Here the dipole mode is significantly broadened, so much that the amplitude decreases whilst redshifting. Conversely the quadrupole resonance is redshifted and increases in amplitude because there is no significant broadening.
\subsubsection{Substrate}

\begin{figure}[t]
\centering
\makebox[\textwidth][c]{\includegraphics*[width=1.5\textwidth]{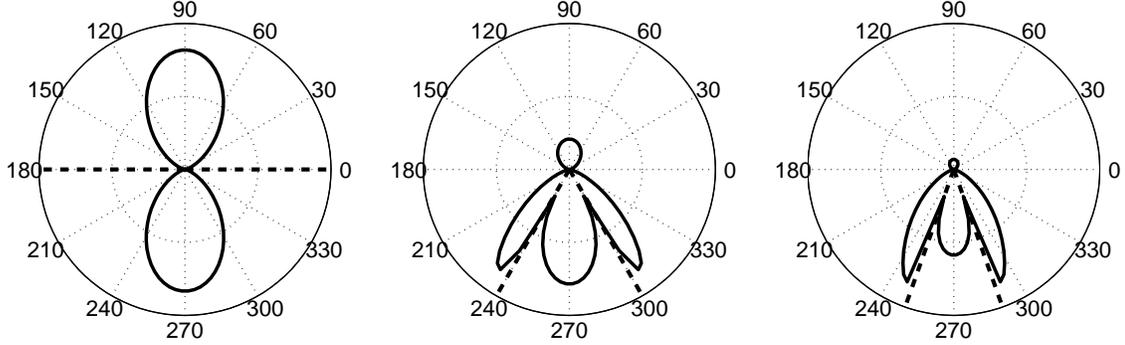}}
\caption[]{The angular scattered power (\textit{solid lines}) from hemispherical Ag nanoparticles at a vacuum / substrate interface for different substrate refractive indices. The lines subtending the angle of total internal reflection inside the substrate (\textit{dashed lines}). The substrate indices are (\textbf{left}) $n = 1.0$, (\textbf{middle}) $n= 2.0$, (\textbf{right}) $n= 3.0$, all substrates are non absorbing. The light is incident from above and images are in the plane of polarization}
\label{Angular_Distribution}    
\end{figure}

The more interesting case is that of a particle on a substrate\index{Substrate} or at a layer interface\index{Layer Interface}. Although we see, broadly speaking, similar redshifting behavior to the homogeneous medium case there are some notable exceptions. Resonances can couple to just one particle-medium interface, for example we take the upper half space as vacuum and only change the refractive index of the lower half space, i.e. a particle on a substrate. Then resonances coupled to the air-particle interface will not be redshifted at all by the changing substrate. Through controlling the amount of surface contact of the particle to each layer at a layered interface, different resonances can be shifted at different rates. Perhaps more importantly for optoelectronic applications is the effect on the near field and far field radiation. Both of these radiation types will become more localized to the material with a higher refractive index, due to the higher density of optical states\index{Optical Density of States} there. This means it is possible to have directional radiation from the particles. Figure \ref{Angular_Distribution} shows the angular distribution of the far field radiation for silver hemispherical particles at a vacuum / substrate interface for different substrate refractive index values, light is incident from above and the images are in the plane of polarization. In Fig. \ref{Angular_Distribution}(a) there is no real interface as the lower half space is also vacuum, appropriately we see the classic dipole radiation with equal forward and backscattering. The dotted lines show the angle of total internal reflection inside the substrate assuming transmission into vacuum. In this case the dotted lines extend horizontally out to infinity. This is because for light to be trapped inside a ``substrate'' which is the same as the surrounding medium, it would have to propagate horizontally as it cannot be totally internally reflected. In the case of (b) and (c) these lines are not horizontal and indicate the direction in which light will be totally internally reflected, i.e. trapped inside the substrate. Additionally we can see that the forward scattering\index{Forward Scattering} becomes increasingly preferential over the back scattering. The forward scattering is divided into the part which lies inside the cone made by the dotted lines of total internal reflection\index{Total Internal Reflection} and those which lie outside. The scattered light inside this cone will not be totally internally reflected and will be transmitted out of the substrate again if the substrate is non absorbing. The two lobes either side, which lie outside the cone, will be reflected upon attempting to leave the substrate, thus providing light trapping within the substrate. As we can see, the fraction of light which is trapped in this way will increase as the substrate index increases.


\section{Designing Plasmonic Solar Cells}
In the previous two sections we have seen how we can manipulate the plasmonic resonance to change the lineshape and peak resonance position. For all applications of plasmonics to technology it is required to tailor the resonance to the optical system in question. As an example we can take the case of nanoparticles integrated into photovoltaics.

\begin{figure}
\centering
\includegraphics*[width=.7\textwidth]{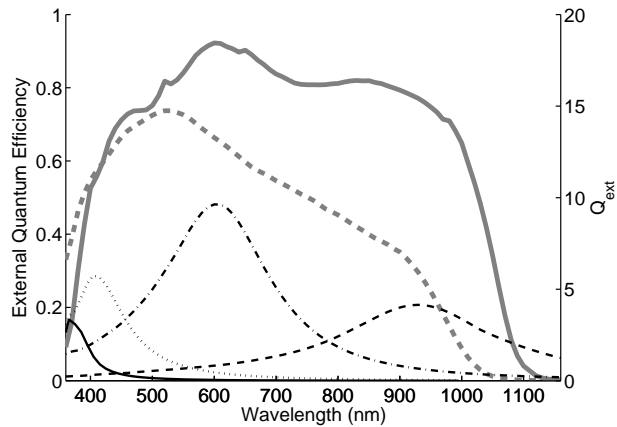}
\caption[]{External quantum efficiency for two different thicknesses of CIGS type solar cells. (\textit{Grey thick solid line}) is for a standard 2$\mu$m thick absorbing layer whereas (\textit{grey thick dashed line}) is for a 450\,nm thick layer, consequently the absorption is reduced particularly near the band gap. Black thin lines represent the normalized scattering cross section for different nanoparticles. (\textit{Solid line}) is a 30\,nm radius sphere in vacuum, (\textit{dotted line}) is 50\,nm radius sphere in vacuum, (\textit{dot dashed line}) is short axis 100\,nm long axis 200\,nm prolate ellipsoid in vacuum and the (\textit{dashed line}) is short axis 100\,nm long axis 200\,nm oblate ellipsoid on an aluminium doped zinc oxide surface}
\label{Quantum_Efficiency}    
\end{figure}

\index{Photovoltaic}Photovoltaic solar cells are required to become thinner to meet cost pressures from other energy sources. Upon making the cells thinner the absorption in the cells will reduce because the light path through the solar cell\index{Solar Cell} has become shorter, this can be seen in Fig. \ref{Quantum_Efficiency}. This will particularly affect light at the red end of the spectrum, at energies slightly above the band gap of the semiconductor absorbing layer, where the absorbance of the semiconductor is low. The first step to increasing the efficiency of solar cells is to change the absorbing material to one with a higher absorption coefficient. The search for better semiconducting materials for photovoltaics still continues today. However this search has already been fruitful, semiconductors such as copper indium gallium selenide (CIGSe) offer similar absorption to crystalline silicon for a fraction of the material thickness \cite{CIGSe}. But there is increasing interest in making the thickness of even thin film cells even lower. In this case we are back to the same problem of not enough absorption.

 For light incident perpendicular to the plane of the solar cell, we can increase the path length inside the solar cell by scattering the light into high angles. As we have seen, this can be achieved by placing nanoparticles at an interface between two media with different refractive indices. To minimize the absorption losses in the metal we would like to select a plasmonic material with low losses in the optical region, Sect. \ref{Material_Effects} showed that Ag was a good choice based on this criterion. The other materials used in a photovoltaic device cannot be chosen so easily without compromising the original device design, therefore generally one seeks to place the particles at a preexisting interface within the layered system compromising the photovolatic. For simplicity we choose the top interface between air and aluminium doped zinc oxide.

As previously discussed we require an absorption enhancement at energies close to the semiconductor band gap. For a CIGS cell that would be at energies just above 1.1\,eV or wavelengths shorter than 1100\,nm. Therefore we would like to have a strong scattering response from our nanoparticles in this region. For small spherical silver particles the resonance will typically be at much shorter wavelengths. To achieve device functionality it is necessary to change the size and shape of the particles, using the principles developed in Sect. \ref{Shape_Effects}. We wish to redshift the spectrum to the correct position, broaden it to cover a large part of the region of absorption loss (here 500-1100\,nm) and have a strong scattering response. Starting from the small nanoparticles we can increase the radius until the dipole scattering response, $Q_{sca}$, is maximised. Then we elongate the particle to redshift and broaden the resonance. Finally taking into account the effect of the substrate gives us the final resonance position shown. The scattering will have a significant high angle component allowing for a light path enhancement inside the absorbing layer.  Due to the extreme broadening the resonance is much less pronounced.  To determine the most beneficial amount of broadening is an optimization problem for which we need a fast, accurate and precise method to evaluate the gain in absorption from the nanoparticles. This is why the finite element method has been employed in this chapter. Table \ref{Summary_Table} provides a summary of the different effects discussed in this chapter.

\begin{landscape}
\begin{table}[!pb]
\centering
\caption[]{Table summarizing the different shape and material parameters for metallic nanoparticles and how they affect observable quantities}
\renewcommand{\arraystretch}{1.2}
\setlength\tabcolsep{5pt}
\begin{tabular}{@{}p{2.75cm}p{6cm}p{4cm}p{4cm}@{}}
\hline\noalign{\smallskip}
 & Normalized Extinction Cross Section & Scattering Efficiency & Angular Distribution \\
\hline\noalign{\smallskip}
Length $\|$ to \textbf{E} & Increase causes redshift and increased intensity & Increase correlates with increase in scattering efficiency& Remains dipolar\\
Length $\|$ to $\vec{k}$ & No redshift, increase in intensity, multipolar modes emerge & No effect& Increased transmission due to multipolar scattering \\
Length $\|$ to \textbf{H}& No redshift, decrease in intensity due to coherence loss& Increase correlates with increase scattering efficiency& Remains dipolar\\
Surface Curvature & Higher curvature causes redshift and increased intensity & No change & Remains dipolar\\
Material & Low loss metals will have a higher intensity & Lower absorption loss increases scattering efficiency& Each material will show multipolar modes for different wavelengths\\
Medium Refractive Index & Increases will redshift the resonance whilst broadening and due to radiative damping & No effect & Multipole orders will become spectrally separated due to redshifting at different rates\\
Substrate & Redshifts will only occur to resonances localized to the side with an increasing refractive index & No effect & Increasing substrate index causes strong forward total scattering and increased high angle scattering\\
\hline
\end{tabular}
\label{Summary_Table}    
\end{table}
\end{landscape}
\section*{Acknowledgements}
The authors would like to thank G. Yin for the experimental data shown in Fig. \ref{Quantum_Efficiency} and P. Andr\"a for the Mie theory results shown in Fig. \ref{Comparison_To_Mie}.  The authors would like to acknowledge funding by the DFG (German Research Foundation) in the DFG research center MATHEON and the funding from the Helmholtz-Association for Young Investigator groups within the Initiative and Networking fund.

\bibliographystyle{plain}
\bibliography{Design_Principles_for_Plasmonic_Nanoparticle_Devices}
%


\printindex
\end{document}